\documentstyle[12pt]{article}
\def\be{\begin{equation}}
\def\ee{\end{equation}}
\def\ba{\begin{eqnarray}}
\def\ea{\end{eqnarray}}
\def\bq{\begin{quote}}
\def\eq{\end{quote}}
\def\PL{{ \it Phys. Lett.} }
\def\PRL{{\it Phys. Rev. Lett.} }
\def\NP{{\it Nucl. Phys.} }
\def\PR{{\it Phys. Rev.} }

\begin{document}

\renewcommand{\theequation}{\arabic{section}.\arabic{equation}}
\newcommand{\dftwo}{(\nabla\phi)^2}
\newcommand{\dstwo}{(\nabla\sigma)^2}
\newcommand{\dffour}{(\nabla\phi)^4}
\newcommand{\riemtwo}{R_{\mu\nu\sigma\rho}R^{\mu\nu\sigma\rho}}
\newcommand{\rhh}{R^{\mu\nu\sigma\rho}
H_{\mu\nu\alpha}H_{\sigma\rho}{}^{\alpha}}
\newcommand{\hfour}{H_{\mu\nu\lambda}H^{\nu}{}_{\rho\alpha}
H^{\rho\sigma\lambda}H_{\sigma}{}^{\mu\alpha}}
\newcommand{\htwohtwo}{H_{\mu\rho\lambda}
H_{\nu}{}^{\rho\lambda}H^{\mu\sigma\alpha}H^{\nu}{}_{\sigma\alpha}}
\newcommand{\dP}{{\dot\phi}}\
\newcommand{\tr}{{\rm Tr}}
\newcommand{\si}{{\sigma}}
\newcommand{\ddt}{\frac{\rm d}{{\rm d} t}}
\newcommand{\pa}{\nabla}
\newcommand{\lz}{\lambda_0}
\newcommand{\ets}{{e^{2\sigma}}}
\newcommand{\emts}{{e^{-2\sigma}}}
\newcommand{\aplz}{\alpha'\lambda_0}

\thispagestyle{empty}
\begin{flushright}
CERN-TH/97-222\\
WATPHYS-TH-97/11\\
hep-th/9708169\\
August 1997
\end{flushright}
\vspace*{1cm}
\begin{center}
{\Large \bf Tailoring T-Duality Beyond the First Loop}
 \\
\vspace*{1cm}
Nemanja Kaloper\footnote{Based on the talks given by N. Kaloper
at the conferences {\it Black Holes: Theory and Mathematical Aspects} at
Banff, AB, Canada and the $7^{th}~C^2GR^2A$ in Calgary, AB, Canada,
May 31 - June 7, 1997.}
\footnote{E-mail: kaloper@hepvs6.physics.mcgill.ca}\\
\vspace*{0.2cm}
{\it Department of Physics, University of Waterloo}\\
{\it Waterloo, ON N2L 3G1, Canada}\\
\vspace*{0.4cm}
Krzysztof A. Meissner\footnote{Permanent address: 
Institute for Theoretical Physics,
Ho{\.z}a 69, 00-689 Warszawa, Poland.} 
\footnote{E-mail: Krzysztof.Meissner@cern.ch} \\
\vspace*{0.2cm}
{\it Theory Division, CERN}\\
{\it 1211 Geneva 23, Switzerland}\\
\vspace{2cm}
ABSTRACT
\end{center}
In this article we review a recent calculation of the
two-loop $\sigma$-model corrections to the $T$-duality map 
in string theory. Using the effective action 
approach, and focusing on backgrounds with 
a single Abelian isometry, we give the 
$O(\alpha')$ modifications of the lowest-order 
duality transformations. The torsion 
plays an important role in the theory to
$O(\alpha')$, because of the Chern-Simons couplings
to the gauge fields that arise via dimensional reduction. 
\vfill
\setcounter{page}{0}
\setcounter{footnote}{0}
\newpage
\section{Introduction}

It is widely believed that 
General Relativity requires certain alterations in order
to be brought into accord with quantum mechanics. The theory
does not yield a satisfactory account of phenomena 
at very small scales even in the context of classical physics. 
As an example, one can take the singularity theorems of
Hawking and Penrose, which show that a gravitating system 
contains singular regions, where the theory breaks down. 
The strongest contender to date for 
the extension of General Relativity into
the quantum realm is string theory. It has brought about the belief 
that matter consists of very small extended objects, 
strings and $p$-branes, the size of which is of the 
order of Planck length. 
The finite size of the elementary building blocks in string
theory could lead to dramatic modifications of small scale
physics, resolving some of the problems faced by General
Relativity.  

In the last few years, we have witnessed 
a rapid and profound development of string theory, 
leading to the establishment of an interconnection of different
string constructions. The principal
tool and guide in the course of this unification 
of string theory was the concept of duality. 
In technical terms, duality arises because of the
considerably richer symmetries of 
string theory than in ordinary General Relativity. In
the string spectrum, in addition to the graviton, there appear
other degrees of freedom, such as the scalar dilaton 
and the antisymmetric torsion tensor 
(or the Kalb-Ramond field), with precisely 
determined couplings. 
String duality symmetries 
arise from the invariance of the theory 
under the exchange of the degrees of
freedom in the string spectrum.
Dualities provide the natural maps between 
seemingly different string theories, and
not only different solutions in the 
phase space of a single theory \cite{witten}).
This has led to the
uniqueness proof of string theory, whereby 
all consistent string constructions 
have been recognized as the facets  
of a single fundamental theory, labeled the M-theory. 

A duality symmetry we have studied in \cite{nkkm} 
was the so-called
$T$-duality, or string scale-factor 
duality. At the level of the 
world-sheet $\sigma$-model, it has been identified by
Buscher \cite{buscher} as a simple Hodge 
duality of a cyclic target coordinate. This 
duality has been also investigated, and generalized, in
\cite{duff}-\cite{mahsch}.
Most of these investigations have been conducted
at the one-loop level of the effective action 
approach to string theory, 
where the action is truncated down to 
the terms of the second order in derivatives. 
However, it has been argued, and in some special
situations proven, that this symmetry
is exact order-by-order 
in perturbation theory \cite{mv12,hassen,ts91,panvel}.
It has also been shown that the 
lowest-order form of the on-shell $T$-duality
map remains unaffected by higher 
order $\alpha'$ corrections when viewed
as a relation between specific conformal field
theories (CFT's) \cite{ekir}, some dual 
solutions in two dimensions \cite{tsey2d} and
some special supersymmetric solutions 
\cite{exact}. In these cases, the proof relied
on special properties of the 
solutions studied - there either existed an
exact, nonperturbative CFT formulation, 
or the solutions were highly symmetric,
which protected them from acquiring quantum corrections. 
A picture which has emerged from these examples 
is that the $T$-duality map can be expanded as 
a perturbative 
series in the inverse string tension $\alpha'$.

Since dualities play such an important role
in the new formulation of string theory,  the 
question of their validity beyond 
the first loop\footnote{We consider here the loop expansion of the
world-sheet $\sigma$-model in the field-theoretic sense, where 
$\alpha'$ is the loop counting variable. From the string theory point 
of view, these corrections are classical, since $\alpha'$ 
measures the effects of the extended structure of strings.} is very
important one. 
A step towards the inclusion of higher-order corrections 
has been taken in \cite{km}, where $T$-duality map has been determined 
to two loops on backgrounds with all but one cyclic coordinates.
Later, in \cite{nkkm} we have generalized this approach to include the gauge
fields which couple to the winding and momentum modes of the string,
and were not considered in \cite{km}. The situations turned out to be
quite a bit more subtle than in the simpler case, because of the
highly nontrivial role of the torsion field.

\section{$O(\alpha'^0)$}

Here we review the one-loop results, in order to show
the explicit lowest-order $T$-duality.
The lowest-order term in the 
effective action of any string theory
truncated to only the model-independent 
zero mass modes is (throughout this paper we use the
string frame with $e^{-2\phi}$ out front, since the symmetry is
most simply realized there)
\begin{equation}
\Gamma^{(0)} = \int d^{d+1}x \sqrt{\bar g}~ e^{-2 \bar \phi}
\left\{\bar R(\bar g)+4 (\bar \nabla \bar \phi)^2 
-\frac{1}{12}\bar H^2\right\}.
\label{actnor}
\end{equation}
Our convention for the signature of the 
metric is $(-,+,\ldots,+)$, the Riemann
curvature is 
$\bar R^{\mu}{}_{\nu\rho\sigma}=\partial_{\rho}
\bar \Gamma^{\mu}{}_{\nu\sigma}-\ldots$,
and the torsion field strength is the 
antisymmetric derivative of the torsion potential:
$\bar H_{\mu\nu\rho}=\nabla_{\mu}\bar B_{\nu\rho}
+{\rm cyclic permutations}$. 
The overbar denotes the quantities 
in the original, $d+1$-dimensional,
frame, before we carry out the Kaluza-Klein reduction.
Note that the definition of the
torsion field strength $\bar H$ encodes the torsion 
potential gauge invariance: if we shift 
the $\bar B$-field according to 
$\bar B \rightarrow \bar B + d \Lambda$, 
where $\Lambda$ is an arbitrary
one-form, the theory remains unchanged. 
As a consequence, the Bianchi
identity for $\bar H$ takes a very simple form: $d\bar H=0$. 

In the presence of a single 
Killing isometry, we can carry out 
the Kaluza-Klein dimensional reduction down to
$d$ dimensions. The most natural 
way to carry out the reduction is to 
ensure that gauge symmetries are manifest at every step 
of the calculation.
The reduced action will feature two additional 
gauge fields coupled
to the graviton, axion and dilaton, and an additional
scalar field, which is the breathing mode
of the cyclic coordinate. 
The reduction ans\"atz for the metric and dilaton, 
$g_{\mu\nu}$ and $\phi$ respectively, is
\begin{eqnarray}
\label{dans}
d{\bar{s}}^2 = {g}_{\mu\nu} dx^{\mu} dx^{\nu} 
+ \exp(2 \sigma)
(dy + V_{\mu} dx^{\mu})^2 ~~~~~~~ 
\phi = \bar \phi - \frac{1}{2} \sigma 
\end{eqnarray}
All the lowering and raising of the 
indices in the rest of this
work will be done with respect to the 
reduced metric $g_{\mu\nu}$.
The vector field $V_{\mu}$ is the 
standard Kaluza-Klein gauge field, which 
couples to the momentum modes of the theory.
The reduction of the axion 
field has to be done with more care because of 
the anomaly which appears in it. 
In the naive decomposition
of the two-form 
$\bar B = (1/2) \bar B_{\mu\nu} dx^{\mu} \wedge dx^{\nu} 
+ W_{\mu} dx^{\mu} \wedge dy$ 
(here $W_{\mu} = \bar B_{\mu y}$ is the other gauge field,
arising from the ``off-diagonal" 
components of the torsion, and which couples to the winding modes),
the space-time components $\bar B_{\mu\nu}$ contribute to the reduced 
torsion, but they are not invariant under the 
translations along $y$. When $y \rightarrow y'= y - \omega(x)$ 
and $V_{\mu} \rightarrow
V'_{\mu} + \partial_{\mu} \omega$ 
(s.t. the cyclic einbein $E=dy + V_{\mu} dx^{\mu}$
is gauge-invariant), we find 
$\bar B_{\mu\nu} \rightarrow \bar B'_{\mu\nu} = 
\bar B_{\mu\nu} + W_{\mu} \partial_{\nu} \omega -
W_{\nu} \partial_{\mu} \omega$. One could try 
$\hat B_{\mu\nu} = \bar B_{\mu\nu} - 
(W_{\mu} V_{\nu} - W_{\nu} V_{\mu})$,
which in fact does not
change by the local translations 
along $y$, but when $W_{\mu} \rightarrow W'_{\mu} = 
W_{\mu} + \partial_{\mu} \lambda_y$, 
$\hat B_{\mu\nu} \rightarrow 
\hat B'_{\mu\nu} = \hat B_{\mu\nu} -
(\partial_{\mu} \lambda_y  V_{\nu} - 
\partial_{\nu} \lambda_y V_{\mu})$.
So the two gauge
symmetries of the reduced theory are not
decoupled, because the reduced torsion potential
cannot be simultaneously invariant under both of them.
The three-form field strength however must be gauge 
invariant, and so following \cite{mahsch}, we can define
the reduced torsion by $B_{\mu\nu} = \bar B_{\mu\nu} - (1/2)
(W_{\mu} V_{\nu} - W_{\nu} V_{\mu})$. Then, the gauge-invariant
field strength can be written as $H=dB - (1/2) WdV - (1/2)VdW$,
using the form notation. This expression is manifestly
$T$-duality invariant to the lowest order,
and hence is the correct stepping stone towards
extending duality to two loops and beyond.
As we will see below, this field strength 
actually appears in the dimensionally
reduced action, and hence 
is indeed the correct choice for the
reduced dynamical variable.

With the ans\"atz (\ref{dans}), we can now carry out the 
reduction of (\ref{actnor}). Here we will skip the details, 
and just give 
the reduced action; 
it is (after dividing $\Gamma^{(0)}_R$ by
the $y$-volume $\int dy$):
\begin{equation}
\Gamma^{(0)}_R = \int d^{d}x \sqrt{g}~ e^{-2\phi}
\left\{R+4\dftwo-\dstwo-\frac14\ets Z
-\frac14\emts T-\frac{1}{12} H^2\right\}.
\label{actl}
\end{equation}
with $B$ and $H$ given by
\begin{equation}
\label{fintor}
B_{\mu\nu} = \bar B_{\mu\nu} 
- \frac{1}{2} \Bigl( W_{\mu} V_{\nu} - 
W_{\nu} V_{\mu} \Bigr)
\end{equation}
and
\begin{equation}
\label{fintorH}
H_{\mu\nu\lambda} = \nabla_{\mu} 
B_{\nu\lambda} 
- \frac{1}{2} W_{\mu\nu} V_{\lambda} 
- \frac{1}{2} V_{\mu\nu} W_{\lambda}
+ cyclic~permutations
\end{equation}
The $T$-duality map is (apart from trivial rescalings) 
the transformation 
$\sigma \leftrightarrow -\sigma$, 
$V_{\mu} \leftrightarrow W_{\mu}$, 
and it is obvious that the action
(\ref{actl}) is invariant under it. 
The equations of motion which are obtained 
from varying the action (\ref{actl}) 
(and are simply related to the string $\beta$-functions,
in this order) are covariant under $T$-duality: the 
$\beta$-functions of the dilaton, reduced metric and torsion 
are symmetric, the $\beta$-function of the modulus is
antisymmetric, and the 
$\beta$-functions of the gauge fields
get interchanged, 
as expected from the world-sheet $\sigma$-model
realization of $T$-duality as a map which exchanges
the momentum and winding modes (the effect of 
$T$-duality on the $\sigma$-model $\beta$ functions
has also been discussed in \cite{haag}).

\section{$O(\alpha'^1)$}

We are now ready to 
discuss the two-loop corrections. The appearance of 
the $O(\alpha')$ corrections in the
effective action can
be changed by finite renormalizations 
of the string world-sheet 
couplings \cite{tm,gs}. 
In a sense, this blurs the 
notion of string theory quantities as functions
of $\alpha'$ - instead of a 
single set of solutions, one ends up with equivalence
classes, specified
by field redefinitions. 
In order to do any 
calculations, one has to adopt a concrete
scheme, thus fixing the form 
of the counterterms in the effective action
and the functional dependence 
on $\alpha'$. 
A scheme may be specified by simultaneously
requiring manifest unitarity in perturbation theory and linear
realization of duality \cite{km} (for the definition of this
scheme, see also \cite{mavmir,jj}). 
For this action, the relationship
between the $\beta$-functions and the functional derivatives 
turns out to be local, to $O(\alpha')$. 
Hence the covariance of the $\beta$-functions and the invariance
of the action in this scheme, to order $O(\alpha')$, are
equivalent. Here we will review the explicit form of the 
corrections. The effective action to two loops is
\begin{eqnarray}
\label{twola}
\Gamma&=&\int d^{d+1}x \sqrt{\bar g}e^{-2\bar \phi}
\left\{\bar R(\bar g)+4(\bar \nabla \bar \phi)^2 
-\frac1{12}\bar H^2\right.
\nonumber\\
&&+\alpha'\lz\left[
-\bar R^2_{GB}+16\left(\bar R^{\mu\nu}-
\frac12\bar g^{\mu\nu}\bar R\right)\bar \nabla_{\mu}\bar \phi
\bar \nabla_{\nu}\bar \phi 
-16\bar \nabla^2\bar \phi (\bar \nabla \bar \phi)^2
+16(\bar \nabla \bar \phi)^4\right.\nonumber\\
&&+\frac12\left(\bar R_{\mu\nu\lambda\rho} 
\bar H^{\mu\nu\alpha} \bar H^{\lambda\rho}{}_{\alpha}
-2 \bar R^{\mu\nu}\bar H^2_{\mu\nu}
+\frac13 \bar R \bar H^2\right)
-2\left(\bar \nabla^{\mu}\bar \nabla^{\nu}
\bar \phi \bar H^2_{\mu\nu}
-\frac13 \bar \nabla^2 \bar \phi
\bar H^2\right)\nonumber\\
&&\left.\left. -\frac23 \bar H^2 
(\bar \nabla \bar \phi)^2
-\frac1{24} \bar H_{\mu\nu\lambda} \bar H^{\nu}{}_{\rho\alpha} 
\bar H^{\rho\sigma\lambda} \bar H_{\sigma}{}^{\mu\alpha}  +
\frac18 \bar H^2_{\mu\nu} \bar H^2{}^{\mu\nu}
-\frac1{144}(\bar H^2)^2\right]\right\}.
\end{eqnarray}
where we introduce a useful shorthand notation:
\begin{equation}
\bar H^2_{\mu\nu}=\bar H_{\mu\alpha\beta}
\bar H_{\nu}{}^{\alpha\beta},
\ \ \ \ {\rm and} 
\ \ \ \ \ \bar H^2=\bar H_{\mu\alpha\beta}
\bar H^{\mu\alpha\beta}
\end{equation}
The action (\ref{twola}) is identical to the action obtained
by Jack and Jones \cite{jj}, modulo a boundary term.
The parameter $\lambda_0$ allows us to move between different
string theories: $\lambda_0 = -1/8 $ 
for heterotic, $-1/4$ for bosonic, and $0$ for 
superstring. In this sense, our 
calculations are completely general (although, 
they are of course trivial in the case of the superstring 
- where the $O(\alpha')$ terms vanish identically 
and hence the lowest-order $T$-duality does not acquire any corrections;
curiously, in this context $T$-duality is not a map between the 
solutions of a theory but instead a map between different theories).
The $\bar R-\bar H$ terms also include the 
Lorentz-Chern-Simons terms which
emerge in the heterotic theory to two loops.
The curvature squared terms are 
arranged in the Gauss-Bonnet combination,
$\bar R^2_{GB} = \bar R^2_{\mu\nu\lambda\sigma} 
- 4 \bar R^2_{\mu\nu} + \bar R^2$.
While the dimensional reduction of this action may appear 
difficult at first, it is manageable when carried out 
on the tangent space. 
The details of the dimensional reduction can be found in
\cite{nkkm}. The contributions to the 
reduced action contain terms which are invariant 
under the one-loop level duality transformations
$\sigma \leftrightarrow -\sigma$, 
$V_\mu \leftrightarrow W_\mu$ and also terms which 
are not symmetric under this map. It 
is these latter terms which we are interested in
here. They are the ones forcing us to correct 
the one-loop duality map with $O(\alpha')$
contributions. In order to separate the
two-loop contributions into 
one-loop duality-invariant
and duality-noninvariant parts, we can 
apply the transformations,
and work out $\Gamma_{inv} = 
(1/2)(\Gamma_2 + T \Gamma_2)$ and
$\Gamma_{ninv} = (1/2)(\Gamma_2 - T\Gamma_2)$. 
Using this and the results of \cite{nkkm}, 
the duality-violating sector of the 
reduced $O(\alpha')$ action is 
\begin{eqnarray}
\label{noninvtact}
\Gamma^{(2)}_{ninv}&=&\aplz\int d^d x\sqrt{g}e^{-2\phi}\Bigl\{
-4 \nabla_{\mu} \sigma \nabla^{\mu} (\nabla \sigma)^2 
-\nabla_\mu \sigma \nabla^\mu  
\left[e^{-2\sigma} T +e^{2\sigma} Z\right] 
\nonumber \\
&&+ \frac{1}{8} \left[e^{-4\sigma} T^2 - e^{4\sigma} Z^2\right] 
+\frac14 H_{\alpha\beta\sigma} H_{\mu\nu}{}^{\sigma} 
\left[W^{\alpha\beta}W^{\mu\nu}\emts
-V^{\alpha\beta}V^{\mu\nu}\ets \right]  \\
&&+\left[\nabla^\mu V^{\nu\lambda} W_{\mu}{}^{\sigma} -
\nabla^\mu W^{\nu\lambda} V_{\mu}{}^{\sigma} \right] 
H_{\nu\lambda\sigma}
+\frac12 \left[V^{\nu\lambda} W^{\mu\sigma} -
W^{\nu\lambda} V^{\mu\sigma} \right] 
\nabla_\lambda H_{\nu\mu\sigma} \nonumber \\
&&+2 \nabla_\gamma \sigma V_{\mu\nu} W^{\gamma}{}_{\lambda} 
H^{\mu\nu\lambda}
+2 \nabla_\gamma \sigma W_{\mu\nu} V^{\gamma}{}_{\lambda} 
H^{\mu\nu\lambda} \Bigr\} \nonumber
\end{eqnarray}

Given the noninvariant terms (\ref{noninvtact}) 
and the one-loop level duality
$\sigma \leftrightarrow - \sigma, 
V_{\mu} \leftrightarrow W_{\mu}$,
the natural way to reconcile 
them is to interpret
(\ref{noninvtact}) as the $O(\alpha')$ 
terms in the expansion of the
exact $T$-duality map, which presumably 
exists in a complete 
quantum theory, which admits various 
string theories as its limits.
Hence, one ought to be able 
to incorporate these terms by 
redefining the duality-invariant 
one-loop action - in effect shifting
the one-loop level fields by 
amounts proportional to $\alpha'$,
while preserving any 
other symmetries the one-loop level
theory has. With this in mind, 
we only need to ensure that the
noninvariant terms (\ref{noninvtact}) 
be absorbed to $O(\alpha')$.
Any deviation away from 
the perturbative form of 
$T$-duality, at any higher loop order,
can be safely ignored from 
the point of view of the effective 
action, in this order. Although this 
may appear limited to 
the two loop level, it conforms
with the idea that $T$-duality is 
perturbatively exact. At higher orders, we
expect that the terms 
violating the two-loop form of duality, that
will emerge from similar calculations,
can be absorbed away
as further corrections of the 
$O(\alpha'^0)$ and $O(\alpha')$
sectors of the action. In fact,
we expect that such analytical cancellations 
will go on {\it ad infinitum}, yielding a
Taylor-series expression for 
the complete $T$-duality map.  

To find the corrections of the $T$-duality map,
we define the shifts of the
one-loop degrees of freedom as follows:
\begin{equation}
\label{shifts}
\sigma \rightarrow 
\hat \sigma + \alpha' \delta \sigma ~~~~
V_\mu \rightarrow 
\hat V_\mu + \alpha' \delta V_\mu  ~~~~
W_\mu \rightarrow 
\hat W_\mu + \alpha' \delta W_\mu ~~~~
 H_{\mu\nu\lambda} \rightarrow 
\hat H_{\mu\nu\lambda} + \alpha' \delta H_{\mu\nu\lambda} 
\end{equation}
Starting from the one-loop 
action (\ref{actl}),
the shifts induce the $O(\alpha')$ correction 
\begin{eqnarray}
\label{shiftact}
\delta \Gamma = 
\alpha' \int d^d x \sqrt{g}e^{-2\phi}&\Bigl\{&
-2\pa_{\mu}\sigma\pa^{\mu}(\delta \sigma) 
- \frac{e^{2\sigma}}{2}[\delta \sigma V_{\mu\nu}^2 
+ V^{\mu\nu} \delta V_{\mu\nu}] \nonumber \\
&&+ 
\frac{e^{-2\sigma}}{2}[\delta \sigma W_{\mu\nu}^2 
- W^{\mu\nu} \delta W_{\mu\nu}] 
- \frac{1}{6} H^{\mu\nu\lambda} 
\delta H_{\mu\nu\lambda}\Bigr\}
\end{eqnarray}
In this formula we have replaced the
shifted degrees of freedom (denoted by a hat in 
(\ref{shifts})) by the original one-loop ones,
in the spirit of the active interpretation of
symmetries.
As we see from the above, the torsion field
needs to be corrected too, although at the
lowest order it is a singlet under
$T$-duality. It may look puzzling 
why it is necessary to correct a
field which is only a 
passive spectator in the arena of
$T$-duality, in order to 
restore the symmetry. Here we 
must remember that the reduced torsion
couples to the gauge fields 
via the mixed Chern-Simons term. 
Since the vectors 
transform nontrivially under duality, 
their corrections 
must play a key role in the
restoration of duality at 
the two-loop level. Their coupling
to the torsion via the Chern-Simons 
terms however also induces 
the corrections of the torsion 
field. Thus the torsion 
assumes the role of a custodian 
field. Its corrections arise because of 
gauge symmetries and are needed 
to restore duality. The form of the
corrections however is determined by the
need to preserve the anomaly terms and
gauge invariance. These two conditions
require that \cite{nkkm}
\begin{equation}
\label{gauginvthatb}
\delta B_{\mu\nu} = 
c W_{\lambda[\mu} V^{\lambda}{}_{\nu]}
+ \delta W_{[\mu} V_{\nu]} 
+ \delta V_{[\mu} W_{\nu]}
\end{equation}
and
\begin{equation}
\label{gauginv3tor}
\delta H_{\mu\nu\lambda} = 
3c \nabla_{[\mu} \bigl(W_{\nu}{}^{\rho}
V_{\lambda]\rho} \bigr) 
- 3 V_{[\mu\nu} \delta W_{\lambda]} -
3 W_{[\mu\nu} \delta V_{\lambda]}
\end{equation}
where $c$ is a constant to be determined by
duality restoration. Note that up to this constant,
the torsion correction is completely determined by the
gauge sector of the theory. Substituting (\ref{gauginv3tor})
into (\ref{shiftact}), and requiring that (\ref{noninvtact})
and (\ref{shiftact}) cancel each other (i.e. that the
$O(\alpha')$ corrections in (\ref{noninvtact}) are absorbed
by (\ref{shiftact})), determines the functional form of the shifts.
We have found the unique explicit form of these
corrections 
\begin{eqnarray}
\label{corrs}
\delta \sigma &=& -2 \lambda_0 \dstwo
-\frac{\lambda_0}4\ets Z
-\frac{\lambda_0}4\emts T\nonumber\\
\delta V_{\alpha} &=& 2\lambda_0 V_{\alpha\rho}\pa^{\rho}\sigma
-\frac{\lambda_0}2 H_{\alpha\beta\gamma}
W^{\beta\gamma}\emts \nonumber \\
\delta W_{\alpha} &=& 2 \lambda_0 W_{\alpha\rho}\pa^{\rho}\sigma
+\frac{\lambda_0}2 H_{\alpha\beta\gamma}
V^{\beta\gamma}\ets
\end{eqnarray}
with the constant $c=-2 \lambda_0$ \cite{nkkm}.
The resulting two terms in the action, the renormalized one-loop
part and the invariant two-loop part (which we did not give here
for brevity's sake, but which can be found in \cite{nkkm}), then are
manifestly invariant under the one-loop level
form of the $T$-duality map $\sigma \leftrightarrow
-\sigma$, $V_{\mu} \leftrightarrow W_{\mu}$, acting
on the corrected fields. If we return to the 
original one-loop level degrees of freedom (unhatted ones
in the equation (\ref{shifts})), and 
interpret the $O(\alpha')$ shifts as the
two-loop corrections to the $T$-duality map,
we can rewrite the new two-loop $T$-duality
map as 
\begin{eqnarray}
\label{corrTd}
\sigma &\rightarrow & -\sigma - 4 \aplz 
(\nabla \sigma)^2 - \frac{\aplz}{2} [\ets Z + \emts T]
\nonumber \\
V_{\mu} &\rightarrow & W_{\mu} -4 \aplz W_{\mu\nu} \nabla^{\nu} \sigma -
\aplz  H_{\mu\nu\lambda}V^{\nu\lambda} \ets \nonumber \\
W_{\mu} &\rightarrow & V_{\mu} -4 \aplz V_{\mu\nu} \nabla^{\nu} \sigma
+ \aplz  H_{\mu\nu\lambda} W^{\nu\lambda} \emts \\
H_{\mu\nu\lambda} &\rightarrow&  H_{\mu\nu\lambda}
-12 \aplz \nabla_{[\mu} \bigl(W_{\nu}{}^{\rho} V_{\lambda]\rho} \bigr)
- 12 \aplz V_{[\mu\nu}W_{\lambda]\rho}\nabla^{\rho} \sigma \nonumber \\
&&- 12 \aplz W_{[\mu\nu}V_{\lambda]\rho}\nabla^{\rho} \sigma 
- 3\aplz \bigl(\ets V^{\rho\sigma} V_{[\mu\nu}
- \emts W^{\rho\sigma} W_{[\mu\nu} \bigr)  H_{\lambda]\rho\sigma} 
\nonumber
\end{eqnarray}
The full reduced action, containing all one-
and two-loop contributions is invariant under (\ref{corrTd}) 
to order $\alpha'$, as one can check by directly applying these
transformation rules. This is our final result.

\section{Conclusion}

In summary, we have presented here the $O(\alpha')$ two-loop 
corrections to the lowest-order $T$-duality map in string
theory. We have started with the effective field theory of the 
model-independent zero mass sector, which included two-loop
corrections in the manifestly unitary form.
Focusing on the string backgrounds with a
single isometry, we have shown that the theory
is invariant under the two-loop corrected $T$-duality map. 
We have arrived at the form of the corrections 
by an iterative reformulation of the $\alpha'$
expansion: those $O(\alpha')$ terms which violate
the one-loop form of duality induce
$O(\alpha')$ corrections in the original duality map. 
The terms linear in $\alpha'$ should be
thought of as the first subleading terms of the Taylor
expansion of the duality map in $\alpha'$. 
One unusual feature of the scheme we have used
is that the BPS solutions of the lowest-order
action do not retain their form when the two-loop
terms are included. As a result, when
our duality corrections are applied to BPS states,
they contain nonvanishing terms to $O(\alpha')$.
While this may sound odd, given the current 
lore \cite{exact}, one should remember 
that while for BPS states there exists a scheme in which 
the classical solutions are exact 
to all orders in the $\alpha'$ expansion, this of course
need not be true in any scheme. Hence, 
those terms among our two-loop
corrections which do not vanish on BPS backgrounds
should be removed by string field redefinitions.
We will not delve on the details here. Suffice it to say that, 
in some sense, these terms behave like
gauge degrees of freedom. 

Another interesting
feature of our calculation is that the torsion
field strength plays a crucial role in 
restoring two-loop duality. This should not come 
as a complete surprise. As has been
pointed out by Maharana and Schwarz, who discovered
the Chern-Simons terms in the definition of the
torsion field strength in the reduced theory \cite{mahsch}, 
the anomaly was essential in rendering the one-loop theory
$T$-duality invariant. This role of the anomaly seems to
persist to two loops, and raises an interesting possibility
that the concepts of the anomaly and $T$-duality 
may somehow be related in the full quantum theory beyond
the effective action limit (M-theory). 
Checking if such a relationship exists demands
a nonperturbative approach, because of 
the complexity of higher-loop counterterms. 

\vspace{1cm}
{\bf Acknowledgements}

K.A.M is grateful to Gabriele Veneziano for
earlier collaboration 
and for discussions on $O(d,d)$ symmetry.
N.K. would like to thank the Theory Division at
CERN, where this work has 
begun, and in particular G. Veneziano for kind 
hospitality. We are also indebted to 
R.R. Khuri and R.C. Myers for helpful conversations and 
comments. This work was supported 
in part by NSERC of Canada.

\end{document}